\documentclass[lnicst,sechang,a4paper]{svmultln}
\usepackage{amssymb}
\usepackage{graphicx}
\newcommand{\be}{\begin{equation}}
\newcommand{\ee}{\end{equation}}
\newcommand{\bra}{\langle}
\newcommand{\ket}{\rangle}
\newcommand{\bea}{\begin{eqnarray}}
\newcommand{\eea}{\end{eqnarray}}
\newcommand{\dis}{\displaystyle}

\usepackage{url}
\urldef{\mailsa}\path|takaishi@hiroshima-u.ac.jp|
\usepackage[pdfpagelabels,hypertexnames=false,breaklinks=true,bookmarksopen=true,bookmarksopenlevel=2]{hyperref}

\begin{document}

\mainmatter  


\title{An Adaptive Markov Chain Monte Carlo Method for GARCH Model}


%
%
\author{Tetsuya Takaishi}%

\institute{Hiroshima University of Economics,\\
731-0192  Hiroshima, Japan\\
\mailsa
}

%
%

\toctitle{Lecture Notes in Business Information Processing}
\tocauthor{Authors' Instructions}
\maketitle

\begin{abstract}
We propose a method to construct a proposal density for the Metropolis-Hastings algorithm 
in Markov Chain Monte Carlo (MCMC) simulations of the GARCH model. 
The proposal density is constructed adaptively by using the data sampled by the MCMC method itself.
It turns out that  autocorrelations between the data generated with our adaptive proposal density
are greatly reduced.  Thus it is concluded that the adaptive construction method is very efficient 
and works well for the MCMC simulations of the GARCH model.
\end{abstract}

\section{Introduction}
It is well known that
financial time series of asset returns
show various interesting properties which can not be
explained from the assumption that the time series obeys the Brownian motion.
Those properties are classified as stylized facts\cite{CONT}.
Some examples of the stylized facts are
(i) fat-tailed distribution of return
(ii) volatility clustering
(iii) slow decay of the autocorrelation time of the absolute returns.
The true dynamics behind the stylized facts is not fully understood.
There are some attempts to make physical models based on spin
dynamics\cite{Iori}-\cite{POTTS} and they are able to capture some of the stylized facts.

In finance volatility is an important quantity to measure risk.
To forecast volatility, 
various empirical models to mimic the properties of the volatility
have been proposed.
In 1982 Engle\cite{ARCH} proposed Autoregressive Conditional Heteroskedasticity (ARCH) model
where the present volatility is assumed to depend on squares of the past observations.
Later Bollerslev\cite{GARCH} proposed Generalized ARCH (GARCH) model which 
includes additional past volatility terms to the present volatility estimate. 

A conventional approach to infer GARCH model parameters is the Maximum Likelihood (ML) estimation
where the GARCH parameters are obtained as the values which maximaize the likelihood function of the GARCH model.
The maximization of the likelihood function can be done by the maximization tool available in computer libraries. 
A practical difficulty of the maximization procedure is that the output results are often sensitive to 
starting values.

An alternative approach, which recently becomes popular, is the Bayesian inference.
Usually the Bayesian inference procedure is performed  by MCMC methods.
There is no unique way to implement MCMC methods.
So far a variety of methods to MCMC procedure have been developed\cite{Bauwens}-\cite{HMC}. 
In a recent survey\cite{ASAI} it is shown that Acceptance-Rejection/Metropolis-Hastings  (AR/MH) algorithm 
works better than other algorithms.
In the AR/MH algorithm the proposal density is assumed to be a multivariate Student's t-distribution and 
its parameters are estimated by the ML technique.
Here we develop a method to determine parameters of a multivariate Student's t-distribution, 
which does not rely on the ML method. 
In our method the proposal density is also assumed to be a multivariate Student's t-distribution
but the parameters are determined by an MCMC simulation.
During the MCMC simulation, the parameters are updated adaptively using the data generated so far.
We test our method using artificial GARCH data and show that the method substantially reduces 
the correlations between the sampled data and works well for GARCH parameter estimations.

\section{GARCH Model}

The GARCH(p,q) model to the time series data $y_t$ is given by
\be
y_t=\sigma_t \epsilon_t ,
\ee
\be
\sigma_t^2  = \omega + \sum_{i=1}^{q}\alpha_i y_{t-i}^2 
+ \sum_{i=1}^{p}\beta_i \sigma_{t-i}^2,
\ee
where $\omega>0$, $\alpha_i>0$ and $\beta_i>0$ to ensure a positive volatility.
Furthermore the stationary condition given by 
\be
\sum_{i=1}^{q}\alpha_i + \sum_{i=1}^{p}\beta_i <1
\ee
is also required.
$\epsilon_t$ is an independent normal error $\sim N(0,1)$.
In many empirical studies it is shown that 
($p=q=1$) GARCH model well captures the properties of the financial time series volatility. 
Thus in this study  we use GARCH(1,1) model for our testbed.
The volatility $\sigma_t^2$ of the GARCH model is now 
written as 
\be 
\sigma_t^2  = \omega + \alpha y_{t-1}^2 + \beta \sigma_{t-1}^2,
\ee
where $\alpha,\beta$ and $\omega$ are the parameters to be estimated.

Let $\theta=(\omega,\alpha,\beta)$ be a parameter set of the GARCH model.
The likelihood function of the GARCH model is written as
\be
L(y|\theta)=\Pi_{i=1}^{n} \frac1{\sqrt{2\pi\sigma_t^2}}\exp\left.(-\frac{y_t^2}{\sigma_t^2}\right.).
\ee
This function plays a central role in ML estimations and also for the Bayesian inference.

\section{Bayesian inference}
In this section we briefly describe the Bayesian inference which 
estimates the GARCH parameters numerically by using the MCMC method. 
From the Bayes' theorem  
the posterior density $\pi(\theta|y)$  
with data $y=(y_1,y_2,\dots,y_n)$ is given by
\be
\pi(\theta|y)\propto L(y|\theta) \pi(\theta),
\ee
where $L(y|\theta)$ is the likelihood function.
$\pi(\theta)$ is the prior density for $\theta$.
The functional form of $\pi(\theta)$ is not known a priori.  
Here we assume that the prior density $\pi(\theta)$ is constant.
$\pi(\theta|y)$ gives a probability distribution of $\theta$ 
when the data $y$ are given. 

With this $\pi(\theta|y)$ values of the parameters are inferred as the expectation values of
$\theta$ given by
\be
\bra {\bf \theta} \ket = \frac1{Z}\int {\bf \theta} \pi(\theta|y) d\theta,
\label{eq:int}
\ee
where 
\be
Z=\int \pi(\theta|y) d\theta.
\ee
$Z$ is a normalization constant irrelevant 
to MCMC estimations.

\subsection{MCMC}
In general the integral of eq.(\ref{eq:int}) can not be performed analytically.
The MCMC technique gives a method to estimate eq.(\ref{eq:int}) numerically.
The basic procedure of the MCMC method is as follows.
First we sample $\theta$ drawn from the probability distribution
$\pi(\theta|y)$. 
Sampling is done by a technique which produces a Markov chain.  
After sampling  some data, 
we obtain the expectation value as an average value over the sampled data 
$\theta^{(i)}=(\theta^{(1)},\dots,\theta^{(k)})$, 
\be
\bra {\bf \theta} \ket = \lim_{k \rightarrow \infty} \frac1k\sum_{i=1}^k \theta^{(i)},
\ee
where 
$k$ is the number of the sampled data.
The statistical error for $k$ independent data 
is proportional to $\frac1{\sqrt{k}}$.
In general, however, the data generated by the MCMC method are
correlated. As a result the statistical error will be proportional to $\sqrt{\frac{2\tau}{k}}$ 
where $\tau$ is the autocorrelation time between the sampled data.
The autocorrelation time depends on the MCMC method we employ.
Thus it is desirable to choose an MCMC method generating data with a small $\tau$.

\subsection{Metropolis-Hastings algorithm}

The most general and simple method to draw values from a given probability distribution is 
the Metropolis method\cite{METRO} or its generalized version, Metropolis-Hastings method\cite{MH}.
Let $P(x)$ is  a probability distribution from which data $x$ shall be sampled.  
First starting from $x$, 
we propose a candidate $x^{\prime}$ which is drawn from a certain probability distribution $g(x^{\prime}|x)$
which we call proposal density.    
Then we accept the candidate $x^{\prime}$ with a probability $P_{MH}(x,x^{\prime})$ 
as the next value of the Markov chain:
\be
P_{MH}(x,x^{\prime})= \min\left[1,\frac{P(x^{\prime})}{P(x)}\frac{g(x|x^\prime )}{g(x^{\prime}|x)}\right].
\label{eq:MH}
\ee
If $x^{\prime}$ is rejected we keep the previous value $x$. 

When $g(x|x^\prime )=g(x^{\prime}|x)$, eq.(\ref{eq:MH}) reduces to the Metropolis accept probability:
\be
P_{Metro}(x,x^{\prime})= \min\left[1,\frac{P(x^{\prime})}{P(x)}\right].
\ee

\section{Adaptive construction of proposal density}
Disadvantages of the MH method are that the candidate drawn as the next value is not always accepted 
and in general the data sampled by the Markov chain are correlated, which results in increasing 
statistical errors.

If the proposal density is close enough to the posterior density 
the acceptance in the MH method can be high.
The posterior density of GARCH parameters often resembles to a Gaussian-like shape. 
Thus one may choose a density similar to a Gaussian distribution as the proposal density.  
Following \cite{WATANABE,ASAI},
in order to cover the tails of the posterior density
we use a (p-dimensional) multivariate Student's t-distribution given by 
\be
g(\theta)=\frac{\Gamma((\nu+p)/2)/\Gamma(\nu/2)}{\det \Sigma^{1/2} (\nu\pi)^{p/2}}
\left[1+\frac{(\theta-M)^t \Sigma^{-1}(\theta-M)}{\nu}\right]^{-(\nu+p)/2},
\label{eq:ST}
\ee
where $\theta$ and $M$ are column vectors,  
\be
\theta=\left[
\begin{array}{c}
\theta_1 \\
\theta_2 \\
\vdots \\
\theta_p
\end{array}
\right],
M=\left[
\begin{array}{c}
M_1 \\
M_2 \\
\vdots \\
M_p
\end{array}
\right],
\ee
and $M_i=E(\theta_i)$.
$\dis \Sigma$ is the covariance matrix defined as
\be
\frac{\nu\Sigma}{\nu-2}=E[(\theta-M)(\theta-M)^t].
\ee
$\nu$ is a parameter to tune the shape of Student's t-distribution. 
When $\nu \rightarrow \infty$ the Student's t-distribution goes to a Gaussian distribution.
At small $\nu$ Student's t-distribution has a fat-tail.

For our GARCH model $p=3$ and $\dis \theta=(\theta_1,\theta_2,\theta_3)=(\alpha,\beta,\omega)$, 
and thus $\Sigma$ is a $3\times3$ matrix.
We determine these unknown parameters $M$ and $\Sigma$ by MCMC simulations.
First we make a short run by the Metropolis algorithm and accumulate some data.
Then we estimate $M$ and $\Sigma$ from the data. 
Note that there is no need to estimate $M$ and $\Sigma$ accurately. 
Second we perform an MH simulation with the proposal density of eq.(\ref{eq:ST}) with the estimated $M$ and $\Sigma$.
After accumulating more data, we recalculate $M$ and $\Sigma$, and update $M$ and $\Sigma$ of eq.(\ref{eq:ST}).
By doing this, we adaptively change the shape of eq.(\ref{eq:ST}) to fit the posterior density.
We call eq.(\ref{eq:ST}) with the estimated $M$ and $\Sigma$ "adaptive proposal density".

\section{Numerical simulations}
In order to test the adaptive construction method 
we use artificial GARCH data generated with a known parameter set 
and try to infer the parameters of the GARCH model from the artificial GARCH data.
The GARCH parameters are set to $\alpha=0.1$, $\beta=0.8$ and $\omega=0.1$. 
Then using these parameters we generated 2000 data.
For this artificial data we perform MCMC simulations by the adaptive construction method.

Implementation of the adaptive construction method is as follows. 
First we start a run by the Metropolis algorithm.
The first 3000 data are discarded as burn-in process or in other words thermalization.
Then we accumulate 1000 data for $M$ and $\Sigma$ estimations.
The estimated $M$ and $\Sigma$ are substituted to $g(\theta)$ of eq.(\ref{eq:ST}).
We re-start a run by the MH algorithm with the proposal density $g(\theta)$. 
Every 1000 update we re-calculate $M$ and $\Sigma$ and update $g(\theta)$.
We accumulate 199000 data for analysis.
To check $\nu$ parameter dependence on the MCMC estimations  we use 
$\nu=(4,6,8,10,12,20)$ and perform the same MCMC simulation for each $\nu$.
Later we find that $\nu$ dependence on the MCMC results is weak.
Therefore the results from $\nu=10$ simulations will be mainly shown.

\begin{table}[h]
  \centering
  \caption{Results of parameters.}
  \label{tab:1}
  {\footnotesize
    \begin{tabular}{clll}
      \hline
        & \multicolumn{1}{c}{$\alpha$} &
      \multicolumn{1}{c}{$\beta$} &
      \multicolumn{1}{c}{$\omega$} \\
      \hline
\hline
   Adaptive ($\nu=10$) & 0.10374 & 0.7789 & 0.11532  \\
   standard deviation\hspace{3mm} & 0.019   & 0.045  & 0.034 \\
   statistical error   & 0.00006 & 0.0002 & 0.00014 \\
   $2\tau$              & $2.3 \pm 0.2$    & $3.0 \pm 0.3$ & $3.4\pm 0.8$     \\
\hline
   Metropolis          & 0.1033 & 0.7797 & 0.1149   \\
   standard deviation  & 0.019  & 0.045  & 0.034 \\
   statistical error   & 0.0005 & 0.0017 & 0.0012 \\
   $2\tau$              & $440\pm 90$\hspace{2mm}   & $900\pm 190$\hspace{2mm}  & $830\pm 170$\hspace{2mm}   \\
      \hline
    \end{tabular}
  }
\end{table}

For comparison we also make a Metropolis simulation
and accumulate 600000 data for analysis.
In this study the Metropolis algorithm is implemented as follows.
We draw a candidate $\theta^\prime$ by 
adding a small random value $\delta \theta$ to the present value $\theta$:
\be
\theta^\prime = \theta + \delta \theta,
\ee
where $\dis \delta \theta= d(r-0.5)$.
$r$ is a uniform random number in $[0,1]$ and 
$d$ is a constant to tune the Metropolis acceptance.
We choose $d$ so that the acceptance becomes greater than $50\%$.

\begin{figure}
\vspace{5mm}
\centering
\includegraphics[height=4.1cm]{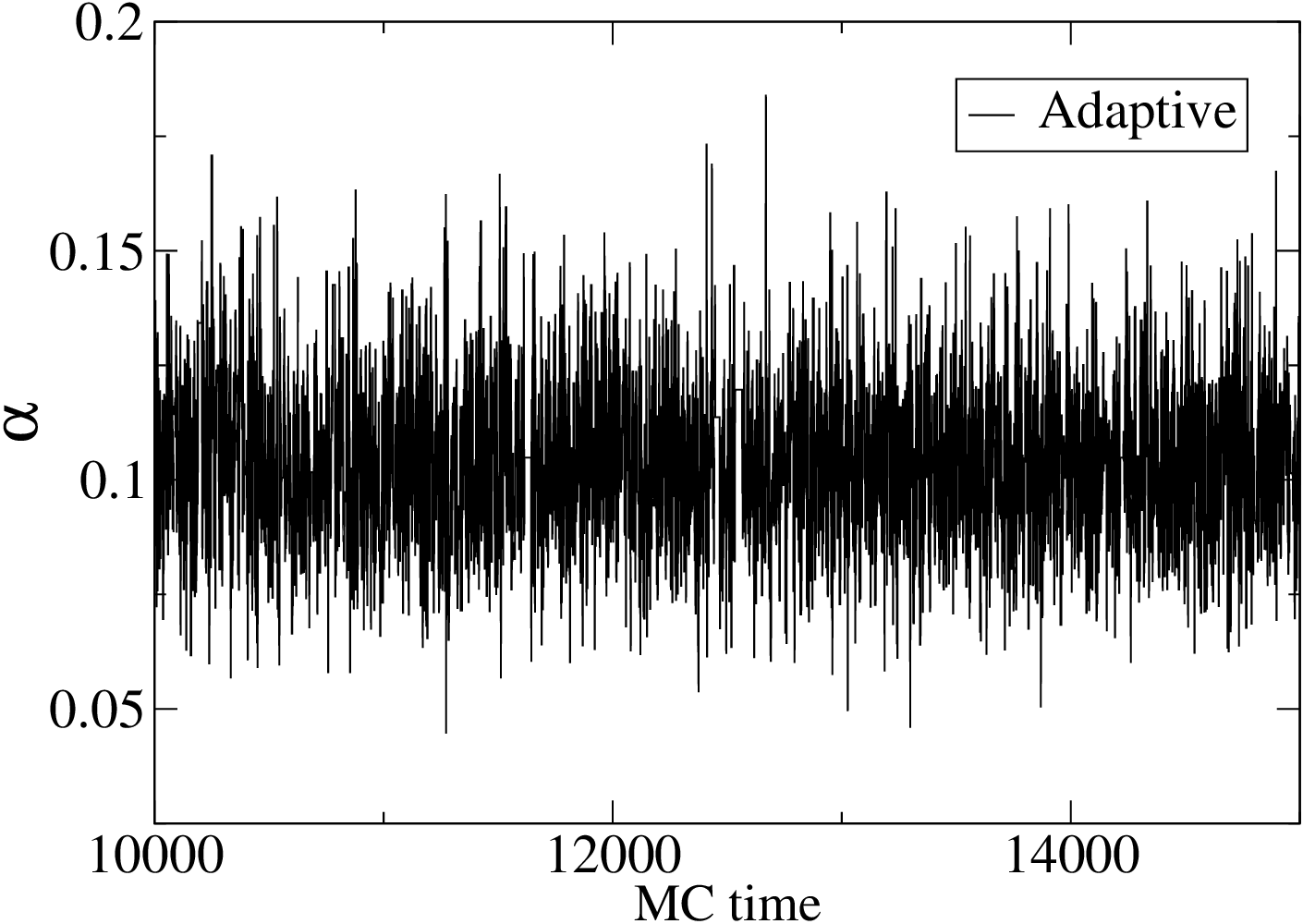}
\includegraphics[height=4.1cm]{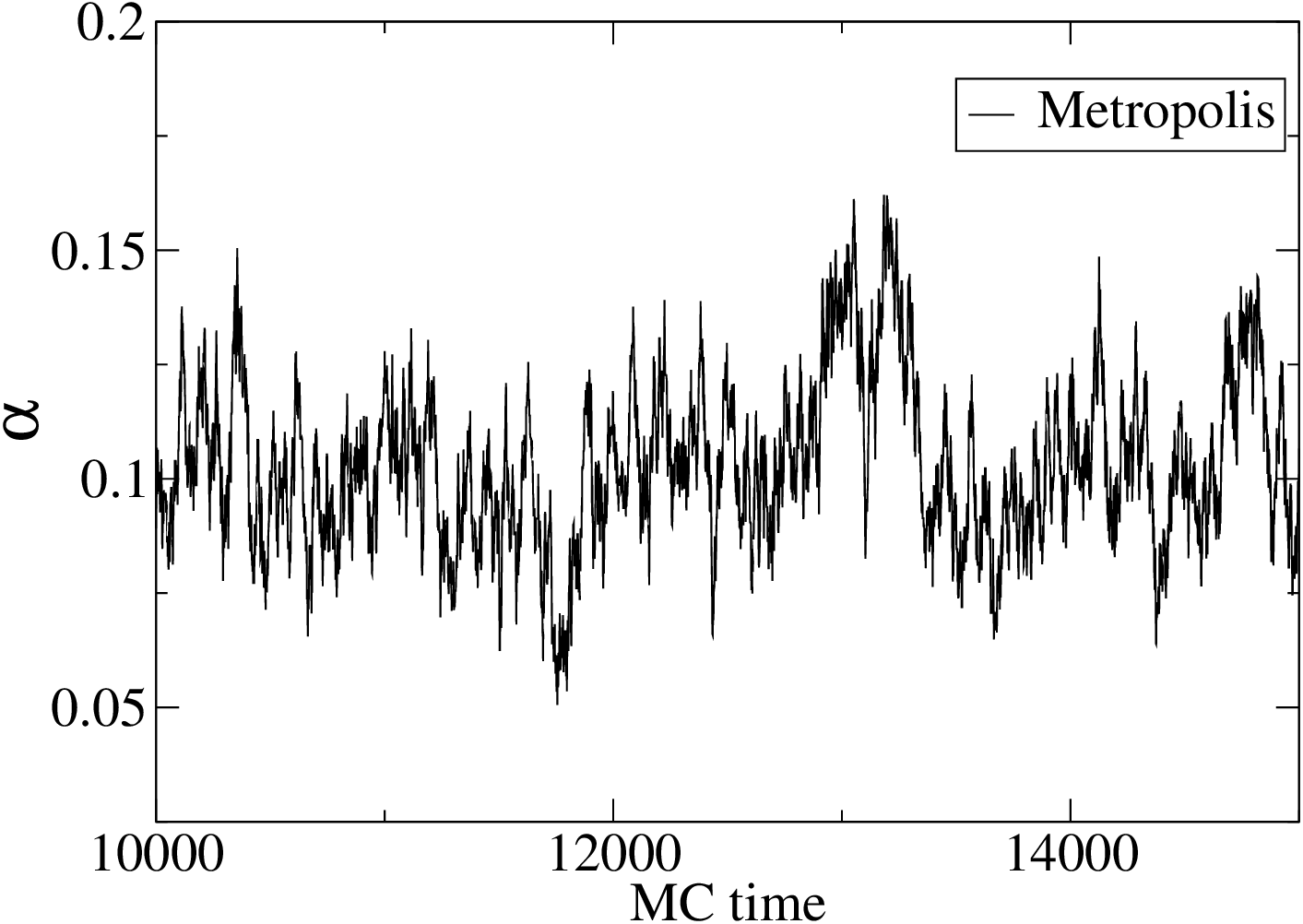}
\caption{
Monte Carlo histories of $\alpha$ from the adaptive construction method with $\nu=10$(left) 
and the Metropolis algorithm(right).
}
\vspace{1mm}
\label{fig:History}
\end{figure}

Fig.~1 compares  the Monte Carlo history of $\alpha$ generated by the adaptive construction method 
with that by the Metropolis algorithm.
It is clearly seen that the data $\alpha$ generated by the Metropolis algorithm are very correlated.
For other parameters $\beta$ and $\omega$ we also see similar behavior.

To quantify the correlation we measure the autocorrelation function (ACF). 
The ACF of certain successive data $x$ is defined by
\be
ACF(t) = \frac{\frac1N\sum_{j=1}^N(x(j)- <x> )(x(j+t)-<x>)}{\sigma^2_x},
\ee
where $<x>$ and $\sigma^2_x$ are the average value and the variance of $x$ respectively.

Fig.~2 shows the ACF for the adaptive construction method and the Metropolis algorithm.
The ACF of the adaptive construction method decreases quickly as Monte Carlo time $t$ increases.
On the other hand the ACF of the Metropolis algorithm decreases very slowly which indicates that
the correlation between the data is very large.

Using the ACF, the autocorrelation time $\tau$ is calculated as
\be
\tau = \frac12 +\sum_{i=1}^{\infty}ACF(i).
\ee
Results of $\tau$ are summarized in Table 1. 
The values of $\tau$ from the Metropolis simulations 
are very large, typically several hundreds.
On the other hand
we see very small correlations, $2\tau \sim 2-3$ for the adaptive construction method.
Thus the adaptive construction method works well for 
reducing correlations between the sampled data.

\begin{figure}
\vspace{5mm}
\centering
\includegraphics[height=4.1cm]{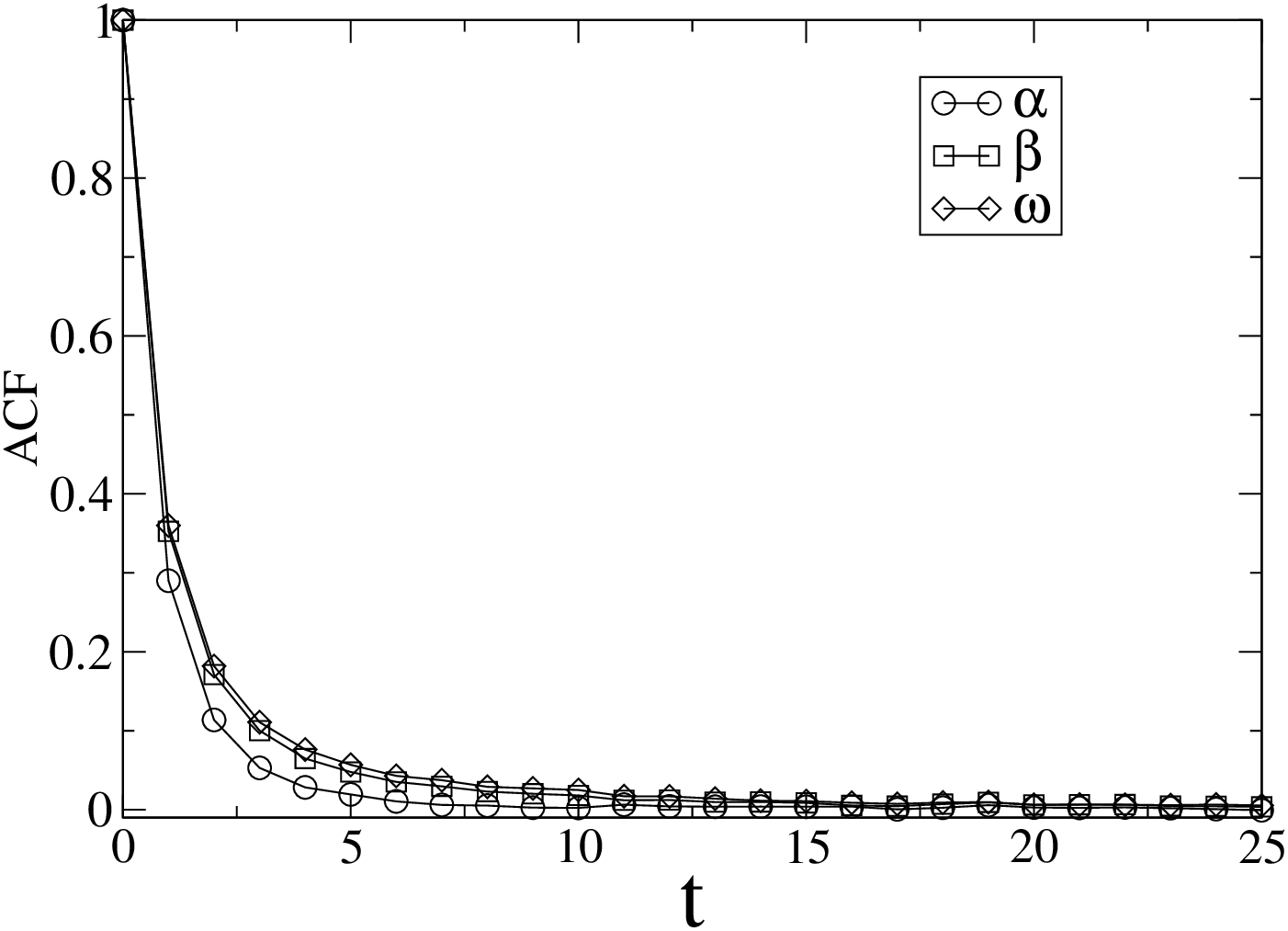}
\includegraphics[height=4.1cm]{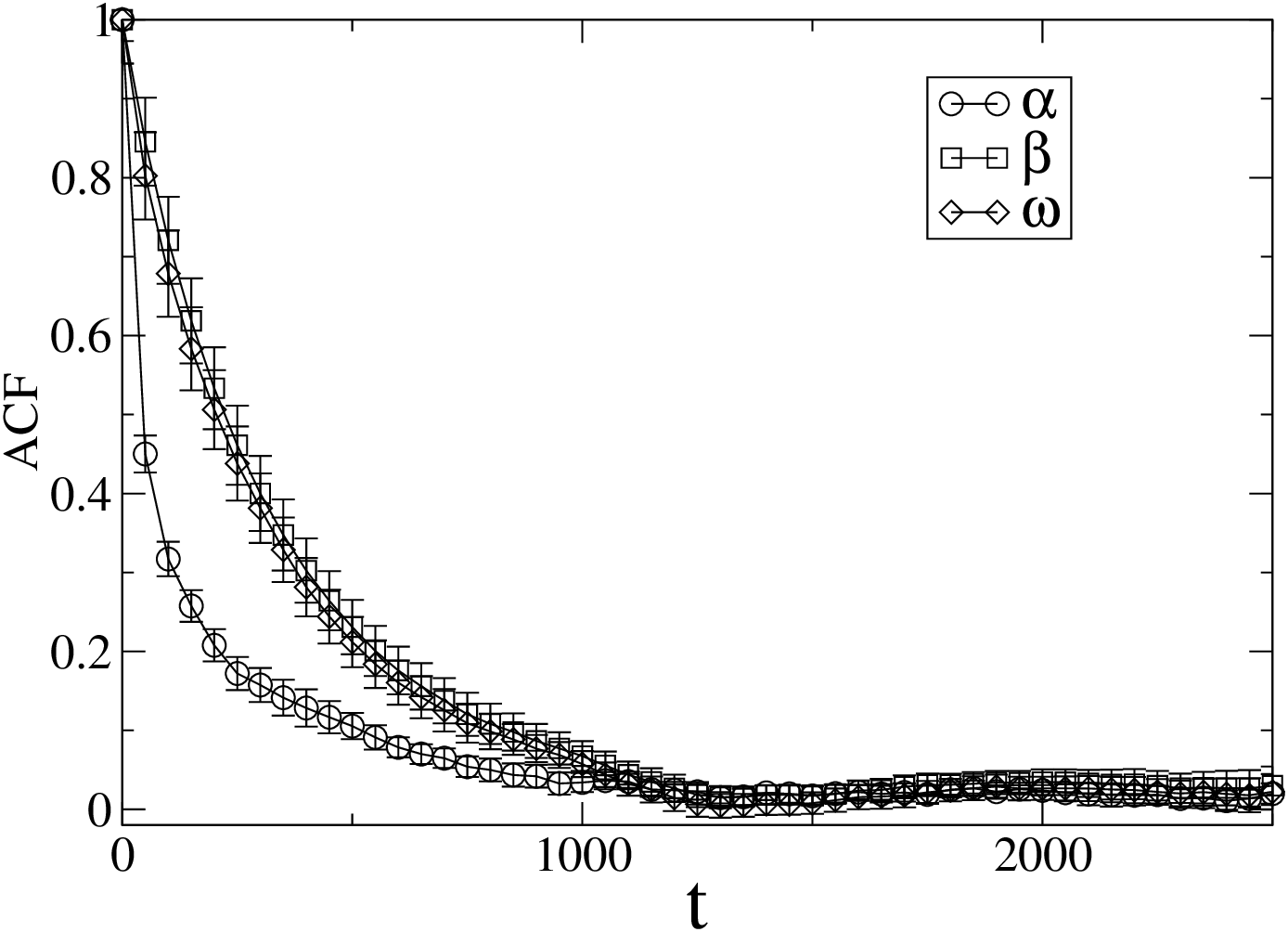}
\caption{
Autocorrelation functions for the adaptive construction method with $\nu=10$ (left) and the Metropolis algorithm (right).
}
\vspace{1mm}
\label{fig:ACF}
\end{figure}

\begin{figure}
\vspace{5mm}
\centering
\includegraphics[height=4.0cm]{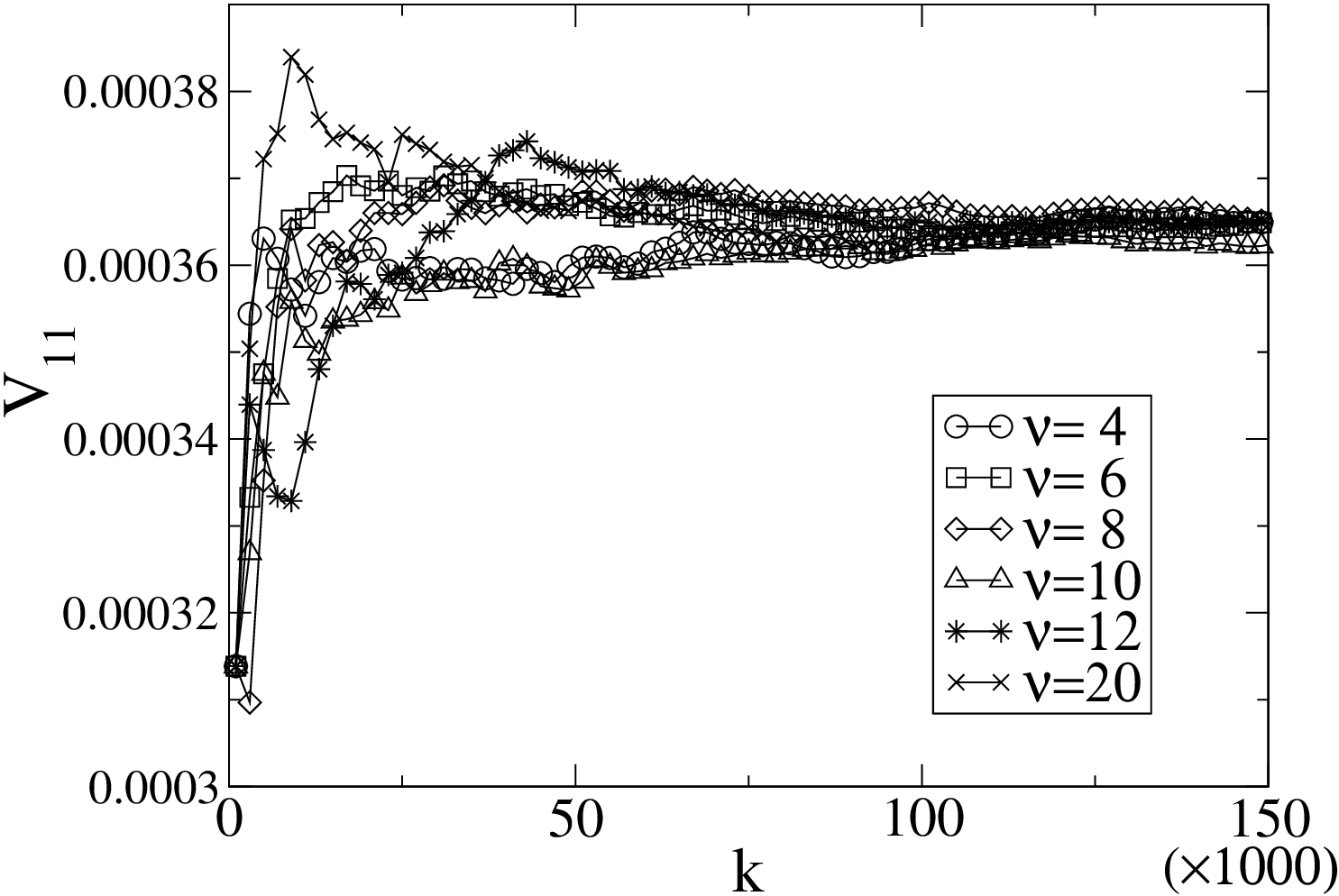}
\hspace{1mm}
\includegraphics[height=4.0cm]{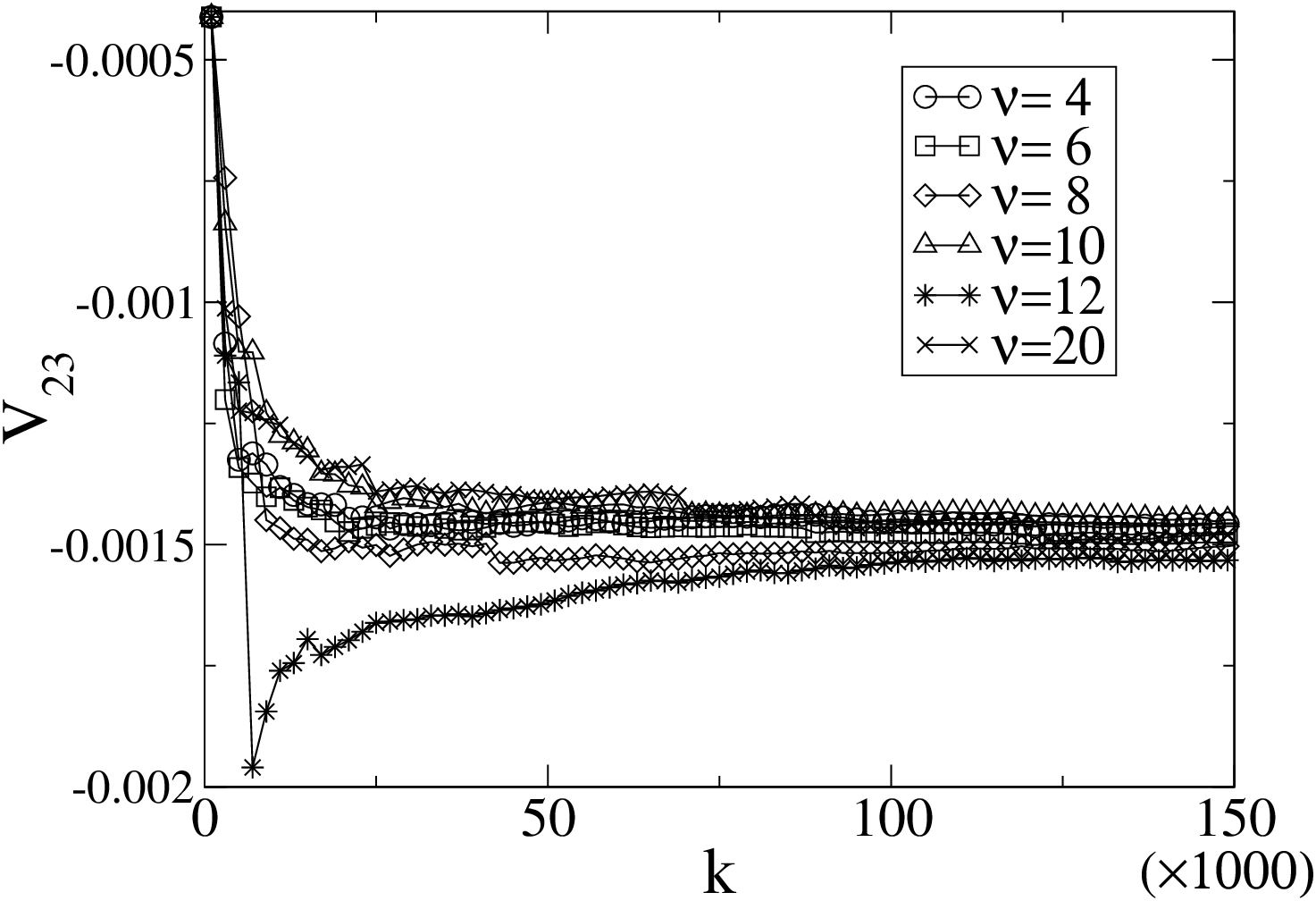}
\caption{
$V_{11}$ and $V_{23}$ as a function of the data size.
}
\vspace{1mm}
\label{fig:SIG}
\end{figure}

We examine how the covariance matrix $\Sigma$ varies 
during the simulations.
Here let us define a symmetric matrix $V$ as 
\be
V=E[(\theta-M)(\theta-M)^t],
\ee 
and $\theta=(\theta_1,\theta_2,\theta_3)=(\alpha,\beta,\omega)$.
Instead of $\Sigma$, we analys this $V$ since $V$ should be same for all $\nu$ 
and it is easy to see the convergence property.

In Fig.~3 we show how $V_{11}(=V_{\alpha\alpha})$ and $V_{23}(=V_{\beta\omega})$ 
change as the simulations are proceeded.
We see that $V_{11}$ and $V_{23}$ converge to some values. 
We also find similar behavior for other $V_{ij}$.
The final output of the matrix elements of $V$ from the simulations is as follows.
\be
V= \left(
\begin{array}{ccc}
 3.6\times 10^{-4}  & -5.8\times 10^{-4} &  2.6\times 10^{-4} \\
-5.8\times 10^{-4}  &  2.1\times 10^{-3} & -1.4\times 10^{-3} \\
 2.6\times 10^{-4}  & -1.4\times 10^{-3} &  1.2\times 10^{-3}
\end{array}
\right).
\ee
From this result we find that $V_{12}(=V_{\alpha\beta})$ and $V_{23}(=V_{\beta\omega})$ are negative, 
and $V_{13}(=V_{\alpha\omega})$ is positive.
Fig.~4 also displays these correlation properties.

\begin{figure}
\vspace{5mm}
\centering
\includegraphics[height=5.6cm]{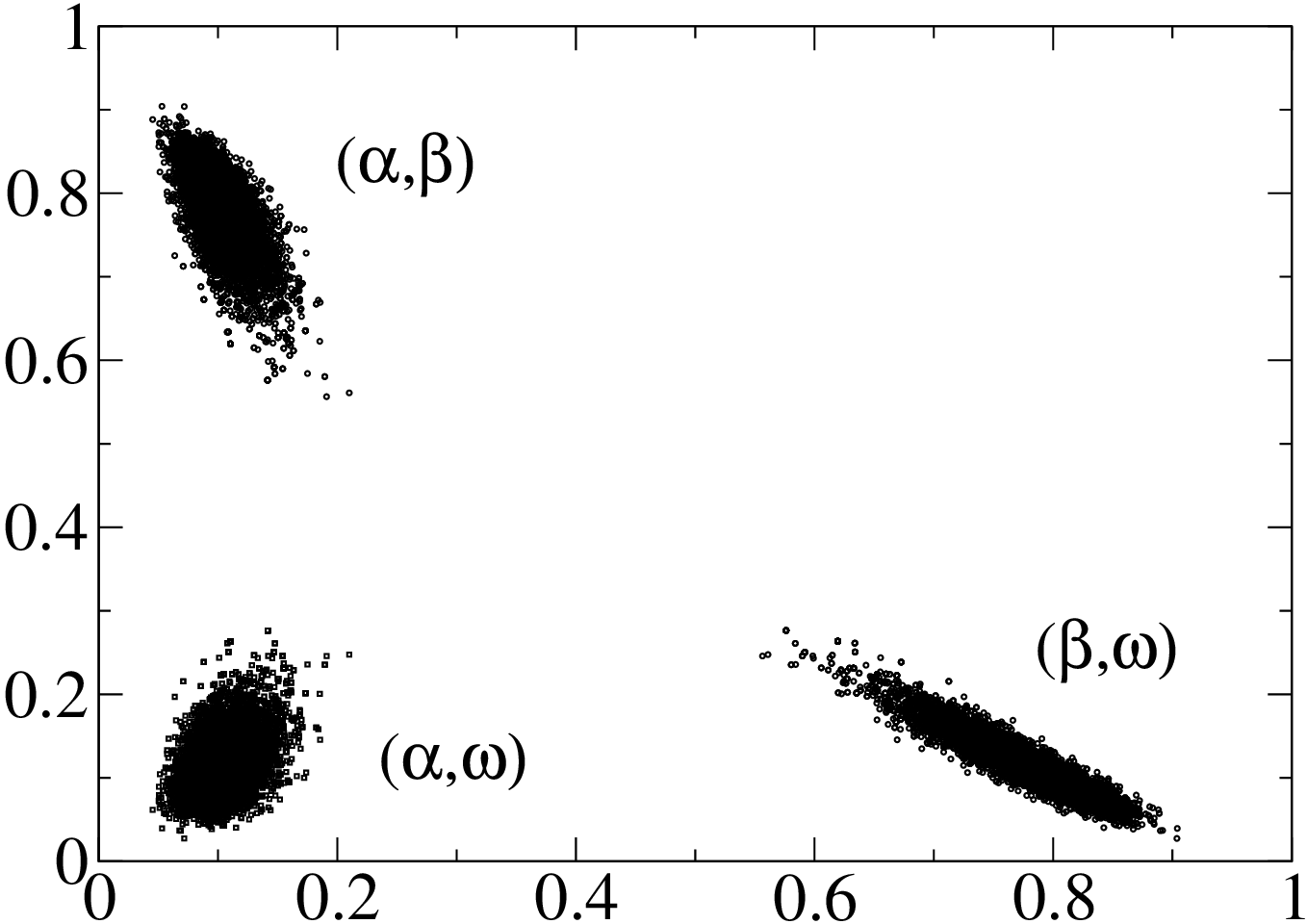}
\vspace{1mm}
\caption{
Scatter plot of sampled data for $(\alpha,\beta)$, $(\beta,\omega)$ and $(\alpha,\omega)$.
}
\vspace{2mm}
\label{fig:Scatter}
\end{figure}

\begin{figure}
\vspace{5mm}
\centering
\includegraphics[height=5.6cm]{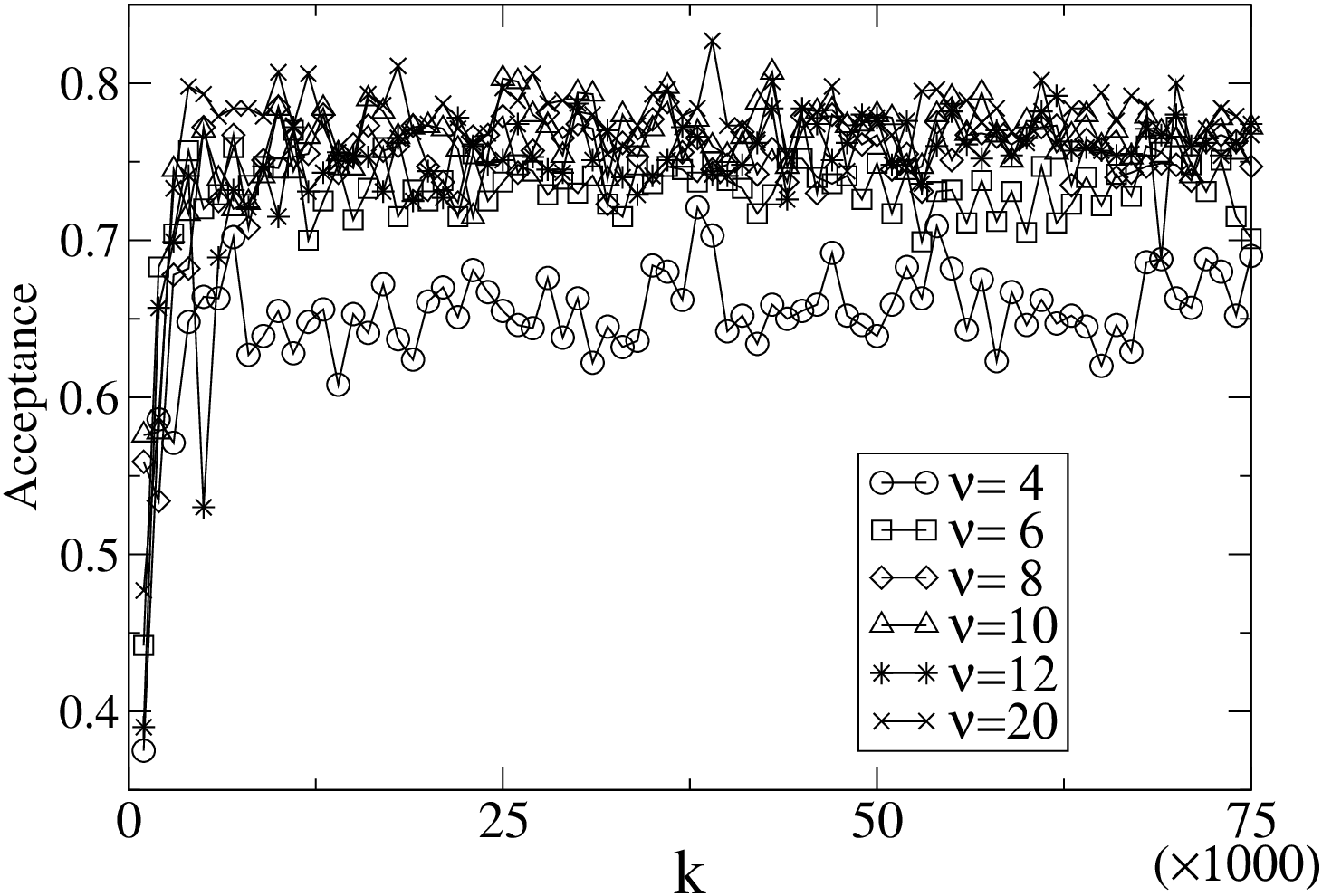}
\vspace{1mm}
\caption{
Acceptance at MH step with the adaptive proposal density.
}
\vspace{2mm}
\label{fig:ACC}
\end{figure}

Fig.~5  shows values of the acceptance at the MH algorithm with the adaptive proposal density of eq.(\ref{eq:ST}).
Each acceptance is calculated every 1000 updates and the calculation of the acceptance is
based on the latest 1000 data.
At the first stage of the simulation
the acceptance is low because $M$ and $\Sigma$ are not calculated accurately as shown in Fig.~3.
However the acceptances increase quickly as the simulations are proceeded
and reaches plateaus.
Typically the acceptances are more than $70\%$ except for $\nu=4$.
Probably $\nu=4$ proposal density is less efficient because 
the tail of the proposal density is too heavy to cover the tail of the posterior density.

\begin{figure}
\vspace{5mm}
\includegraphics[height=4.2cm]{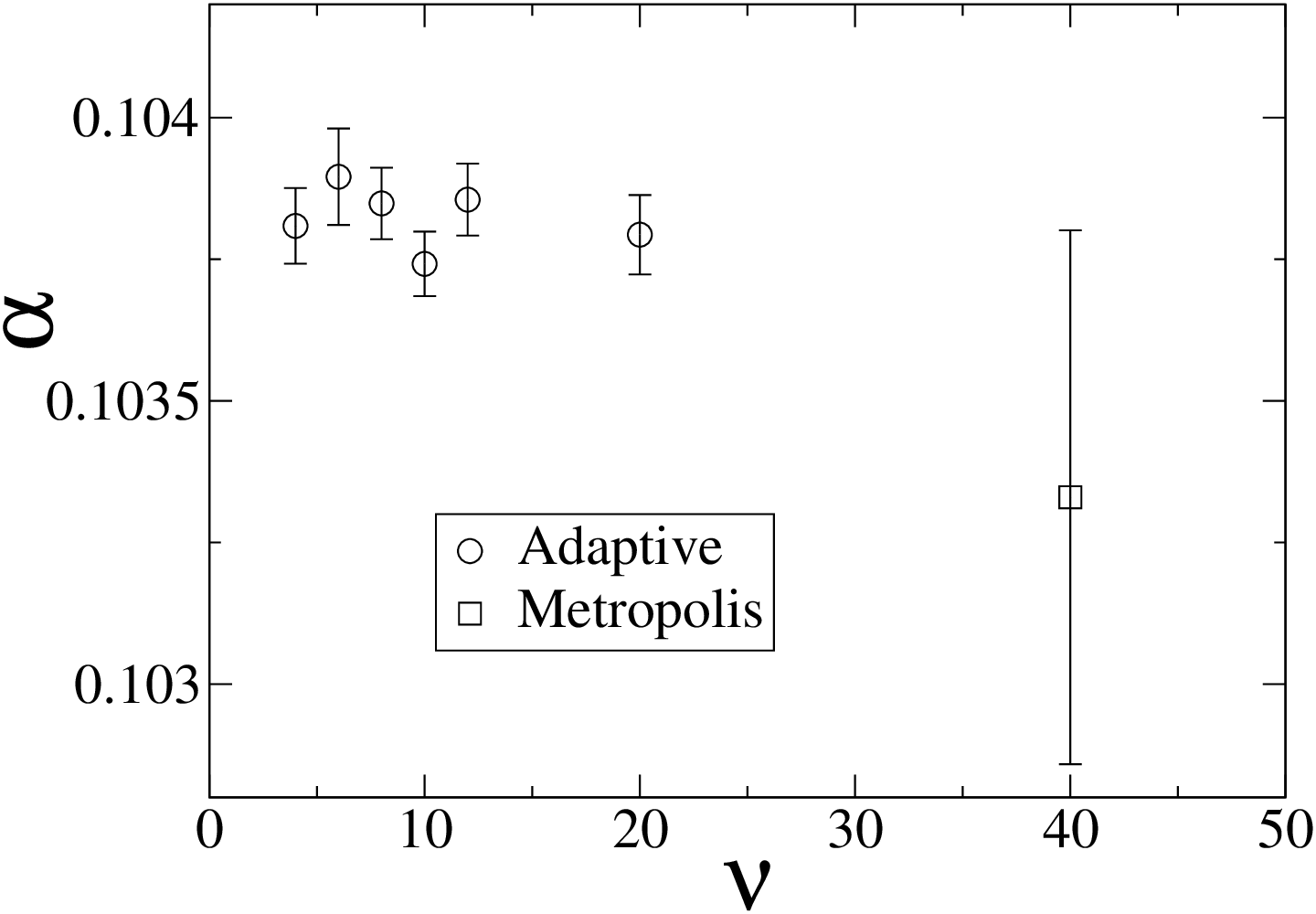}
\vspace{10mm}
\includegraphics[height=4.2cm]{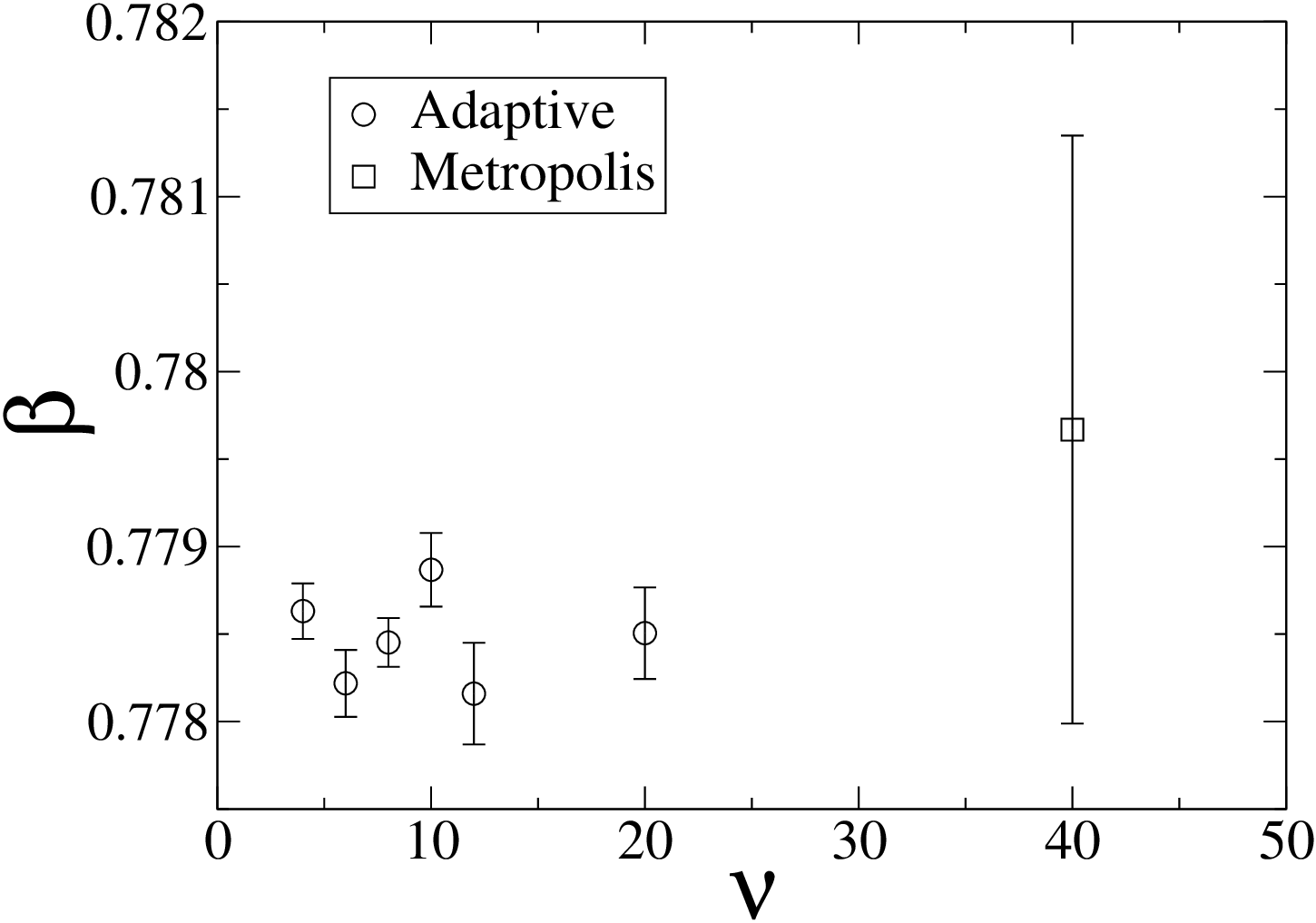}
\includegraphics[height=4.2cm]{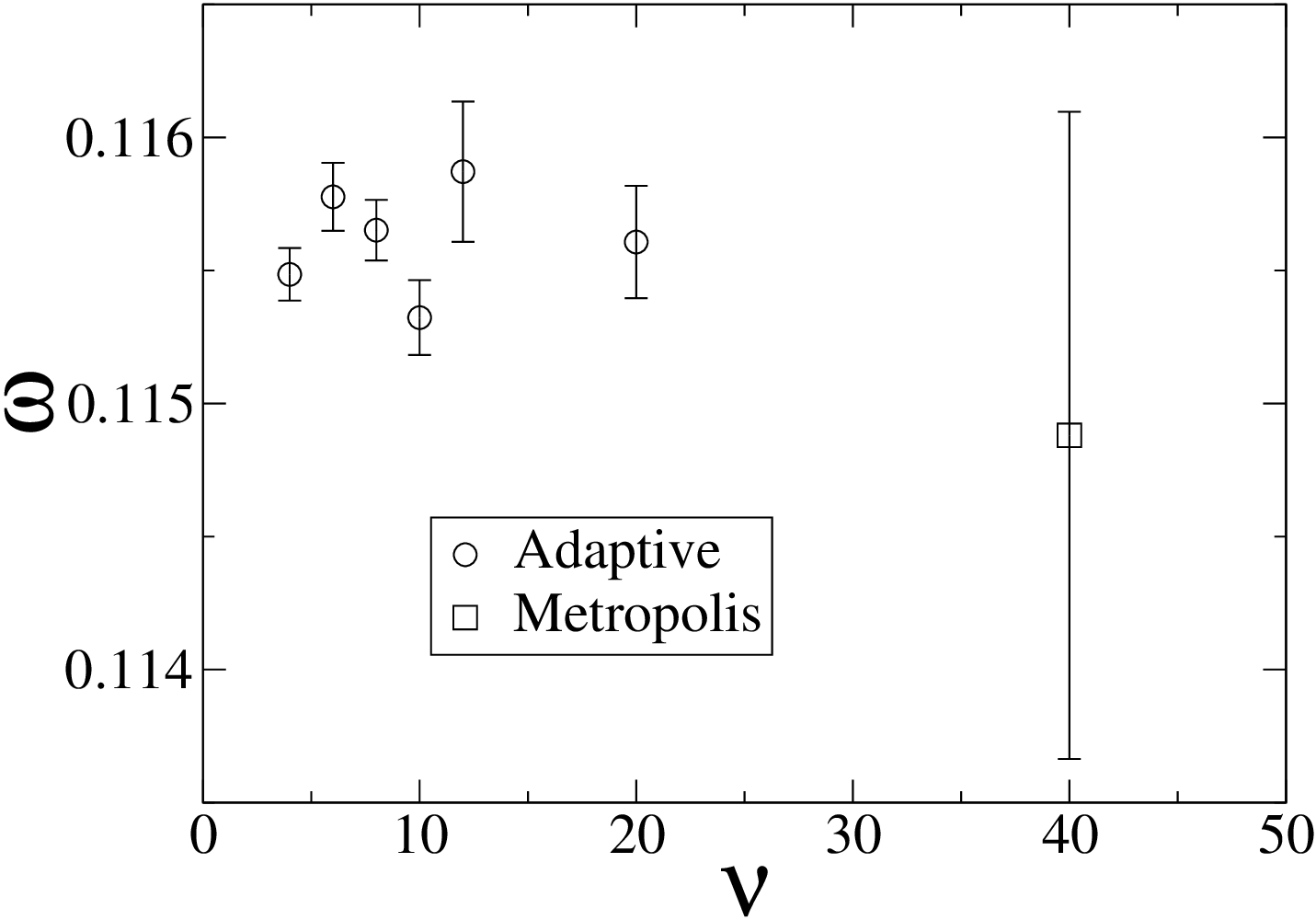}
\caption{
Results of the GARCH parameters estimated by MCMC methods.
The results at $\nu=40$ are from the Metropolis simulations.
The error bars show statistical errors only and they are calculated by the jackknife method.
}
\vspace{2mm}
\label{fig:Omega}
\end{figure}

Fig.~6 shows results of the GARCH parameters estimated by the MCMC methods.
The values of the GARCH parameters are summarized in Table 1.
The results from the adaptive construction method have much smaller error bars and 
are consistent each other. 
On the other hand the Metropolis results have larger error bars
although the number of the sampled data is larger than that of the adaptive construction method.
This is because the data sampled by the Metropolis algorithm are long-correlated. 
The all results with standard deviations agree with the input GARCH parameters ($\alpha=0.1$, $\beta=0.8$ and $\omega=0.1$).
This indecates that the MCMC estimations are correctly done.

\section{Summary}
We proposed a method to construct a proposal density used in the MH algorithm. 
The construction of the proposal density is performed using the data generated by MCMC methods.
During the MCMC simulations the proposal density is updated adaptively.
The numerical results show 
that the adaptive construction method significantly reduces the correlations between the sampled data.
The autocorrelation time of the adaptive construction method is calculated to be $2\tau \sim2-3$. 
This autocorrelation time is similar to that of the AR/MH method\cite{ASAI} which uses the ML estimation.  
Thus the efficiency of the  adaptive construction method is comparable to that of the AR/MH method.
This is not surprising because both methods construct the essentially same proposal density in different ways.
Therefore the adaptive construction method serves as an alternative efficient method for GARCH parameter inference
without using ML estimations.

\section*{Acknowledgments}
The numerical calculations were carried out on Altix at the Institute of Statistical Mathematics
and on SX8 at the Yukawa Institute for Theoretical Physics 
in Kyoto University. 

\end{document}